# *Scopus'*s Source Normalized Impact per Paper (SNIP) *versus* a Journal Impact Factor based on Fractional Counting of Citations

Loet Leydesdorff [a] & Tobias Opthof [b, c]

**Abstract**:

Impact factors (and similar measures such as the Scimago Journal Rankings) suffer from two problems: (*i*) citation behavior varies among fields of science and therefore leads to systematic differences, and (*ii*) there are no statistics to inform us whether differences are significant. The recently introduced SNIP indicator of *Scopus* tries to remedy the first of these two problems, but a number of normalization decisions are involved which makes it impossible to test for significance. Using fractional counting of citations—based on the assumption that impact is proportionate to the number of references in the citing documents—citations can be contextualized at the paper level and aggregated impacts of sets can be tested for their significance. It can be shown that the weighted impact of *Annals of Mathematics* (0.247) is not so much lower than that of *Molecular Cell* (0.386) despite a five-fold difference between their impact factors (2.793 and 13.156, respectively).

[a] Amsterdam School of Communications Research (ASCoR), University of Amsterdam, Kloveniersburgwal 48, 1012 CX Amsterdam, The Netherlands.
[b] Department of Medical Physiology, University Medical Center Utrecht, Utrecht, The Netherlands.
[c] Experimental Cardiology Group, Heart Failure Research Center, Academic Medical Center AMC, Meibergdreef 9, 1105 AZ Amsterdam, The Netherlands.

**Introduction**

Elaborating on ideas first applied to field normalization of the indicators developed by the Center of Science and Technology Studies CWTS at Leiden University (Moed *et al.*, 1995), Moed (2010) has recently proposed a new measure of citation impact for journals called the "source normalized impact per paper" (SNIP) as an alternative to the Journal Impact Factor (Sher & Garfield, 1965; Garfield, 1972) developed by the Institute of Scientific Information (ISI) of Thomson-Reuters. This new indicator—together with the Scimago Journal Ranking (SJR) (Falagas *et al.*, 2008; cf. Leydesdorff, 2009)—is linked at the homepage of Elsevier's *Scopus* database. These indicators can be retrieved directly at http://info.scopus.com/journalmetrics/?url=journalmetrics for all journals included in the *Scopus* database.

In this communication we focus on the SNIP indicator, which tries to solve the problem that citation frequencies in some sciences (e.g., mathematics) are on average significantly lower than in others (e.g., the biomedical sciences) (Leydesdorff, 2008; Waltman & Van Eck, 2009). Moed's (2001) solution is to normalize by using the concept of "citation potential": when authors provide many references, a paper has a higher chance of being cited. Thus, one should normalize in terms of the number of the references in the citing papers. Furthermore, the field-specific citation behavior can be taken into account by studying this on a paper-by-paper basis. For example, if Paper A is cited by Paper B containing $n$ references and by Paper C containing $m$ references, the contributions to the impact of Paper A could be weighted as 1/*n* and 1/*m,* respectively (Narin, 1976; Pinski & Narin, 1976; Small & Sweeney, 1985; Zitt & Small, 2008).

Starting from this idea, the SNIP indicator follows a different path by first aggregating the citing papers at the level of each citing journal. We shall argue that this approach is erroneous because it involves a transgression of the order of operations in mathematics which prescribes that division (i.e., normalization) should precede addition. However, weighting before averaging can provide us with a weighted impact factor that, unlike the ISI-IF or the SNIP, allows for the statistical testing of differences for significance. In other words, the original idea behind the SNIP indicator is a good one, but the elaboration is problematic. Both problems of the IF and its variants—(*i*) the problem of the field-specificity of citation behavior and (*ii*) the lack of statistics—can be solved by using the alternative indicator proposed here.

**The SNIP indicator**

The SNIP indicator is defined as the quotient of two other indicators: (*i*) the journal's *Raw Impact per paper published in the journal* (RIP) divided by (*ii*) the *Relative Database Citation Potential* (RDCP) in the journal's subfield. Note that both the numerator and the denominator are quotients. The numerator, that is, the journal's *Raw Impact per paper published in the journal*, is not essentially different from a three-year Impact Factor. The only difference is technical: in the *Scopus* database this RIP is based on articles, proceedings papers, and reviews both in the numerator and the denominator, while in the ISI database the numerator is provided by any citation to the journal. However, the denominator is delimited similarly in terms of citable items. Citable items include letters in the case of the ISI database. For all practical purposes, however, the numerator of the SNIP indicator can be considered as a three-year IF.

The denominator of the SNIP, that is, the *Relative Database Citation Potential* is defined with reference to the *median* of the *Database Citation Potentials* of the journals in the database. In the *Scopus* database this median value happens to be provided by the *Journal of Electronic Materials.* Moed (2010) compares this journal with *Inventiones Mathematicae* and *Molecular Cell* as examples of journals in fields with low citation density (mathematics) and high citation density (biomedicine), respectively. Citation potential is then defined as the *mean* number of the one- to three-year-old (e.g., 2004-2006) cited references per paper in a citing (2007) journal.

The exercise of the SNIP indicator is complex because normalization is performed both in the numerator and the denominator. In the numerator, the IF-3 is based on averaging skewed distributions (Egghe, 2009; Waltman & Van Eck, 2009), and in the denominator the RDCP is the median of the means.[1] In a different context, Opthof & Leydesdorff (2010; cf. Lundberg, 2007) have shown that the division of means can have significant effects on the rankings, while the proper order of normalizing first (and then taking the average of the normalized values) allows for statistical testing in research evaluation.

---

[1] Since the means can be expected to be distributed normally, one could have considered the mean of these means, or perhaps even better the means of the medians—given the skewedness of the underlying citation distributions (Leydesdorff, 2008; Waltman & Van Eck, 2009).

**Weighting the citation impact in terms of the citing papers (Methods)**

The procedure used for obtaining the relevant papers can be formalized within the *Scopus* database, for example, as follows:

> SRCTITLE("Journal of Electronic Materials") AND (PUBYEAR IS 2004 OR PUBYEAR IS 2005 OR PUBYEAR IS 2006) AND (DOCTYPE(ar) OR DOCTYPE(cp) OR DOCTYPE(re))

Using the button "Cited by," one can retrieve the citing journals and the results can be limited to the publication years and the specific document types. The three journals provided as examples by Moed (2010), however, are also included in the ISI database. We preferred to use this database for pragmatic reasons.

Using the citation search option within the Web of Science, the journal abbreviation can be entered under "cited work" and the citation window 2004-2006 specified. The citable items can be confined in terms of the preferred document types, and the search results, that is, the citing documents, can be confined to publications with 2007 as publication year and also the preferred document types.[2] In order to stay as close as possible to Moed (2010), we used only the three types of documents, and did not include letters, both on the citing and cited sides. The search string in the ISI database can be formulated as follows:

---

[2] There is a difference between publication years and tape years because publications are sometimes late with reference to the calendar year. In this study, we use publication years.

Cited Work=(J Electron Mater or J Electronic Mat* or J Electron Mat* or J Electr Mat* or J Elect Mat or J Elec Mat*)[3] AND Cited Year=(2004-2006) AND Document Type=(Article OR Proceedings Paper OR Review) Timespan=All Years. Databases=SCI-EXPANDED, SSCI, A&HCI.

This results in a retrieval of 2,996 records.[4] The search result can be refined to publication year and document types on the citing side. This leads to a retrieval of 629 documents in this case.

Instead of averaging this for each citing journal, the citation count can be fractionated at the level of the retrieved papers. The *Journal of Electronic Materials*, for example, is cited (629 times) in 151 journals. These citing documents contain 19,459 cited references of which 1,740 to this journal. The references to the *Journal of Electronic Materials* can be weighted as each contributing $1/n$ to the citation impact of the paper ($n$ is the number of references in the citing paper). Note that the journal will sometimes be cited by the same citing paper more than once (for different papers). One can compute accordingly.

These fractions can legitimately be aggregated as fractional counts (Narin, 1976; Small & Sweeney, 1985). In the case of this journal, the weighted citations add up to an impact of 78.926 based on 794 citable items during the period 2004-2006. Thus, the weighted citation impact (per paper) of the *Journal of Electronic Materials* is 78.926 / 794 = 0.099.

---

[3] One of the referees noted that the use of all relevant abbreviations and truncations is necessary for the retrieval since despite the standard list of abbreviations for all journals indexed in the database, various studies have shown that there are also many errors in this list (e.g., Cronin & Meho, 2008).
[4] The user be warned that the citations can only be retrieved in batches of 500 records at a time. This is noted at the end of the page.

| | N citable[5] 2004-2006 | N citing papers 2007,[3] | N cited references[6] | "IF-3 Yrs"[7] | Sum weighted citations | **Weighted impact** | **SNIP 2007** | SJR 2007 | ISI-IF 2007, 2 Yrs |
|---|---|---|---|---|---|---|---|---|---|
| | (a) | (b) | (c) | (d = c/a) | (e) | **(f = e/a)** | **(g)** | (h) | (i) |
| 1. *Invent Math* | 204 | 355 | 740 | 3.627 | 34.479 | **0.169** | **3.294** | 0.065 | 1.664 |
| 2. *Mol Cell* | 923 | 8,038 | 19,058 | 20.648 | 355.958 | **0.386** | **3.696** | 7.110 | 13.156 |
| 3. *J Electron Mater* | 794 | 629 | 1,740 | 2.191 | 78.926 | **0.099** | **1.319** | 0.113 | 1.320 |
| 4. *Math Res Lett* | 221 | 150 | 189 | 0.855 | 8.912 | **0.040** | **1.179** | 0.041 | 0.702 |
| 5. *Ann Math* | 165 | 512 | 998 | 6.048 | 40.720 | **0.247** | **4.979** | 0.104 | 2.739 |

**Table 1**: Various indicators for the five journals

Table 1 provides the results for the three journals studied by Moed (2010). Two more journals in the field of mathematics were added in order to make within-field comparisons possible (see below). The rank order among the journals is not changed by using weighted impact when compared with either the ISI-IF or the three-year impact factor. However, the rank order for the SNIP is different for *Annals of Mathematics* and *Molecular Cell.* This confirms the above claim that changes in the order of operations may have unpredictable effects on the rank order (Opthof & Leydesdorff, 2010; cf. Van Raan *et al.*, 2010; Leydesdorff & Opthof, in preparation).

While the SNIP indicates a value for the *Annals of Mathematics* larger than *Molecular Cell,* the latter still scores higher than the former using the weighted impact, but not even twice as high. The five-fold difference in the ISI impact factors between these two journals is largely due to the size of the reference lists which is much larger in biomedicine than in mathematics. In other words, this indicator convincingly solves the problem of normalizing for different citation behavior.

---
[5] Citable items were in accordance with Moed (2010) defined as only articles, reviews, and conference proceedings papers. All downloads were done on March 29, 2010.
[6] Using the ISI database at the Web-of-Science, one is not able to distinguish whether references are to source or non-source items. The numbers of cited references may therefore be overestimated and weighted impact accordingly could be higher if one would include only ISI-source items, that is, items from journals included on the ISI source list.
[7] We placed quotation marks around this impact factor because it is based on searches in which "letters" were systematically not included.

Correlation analysis of the indicators in this very small set of only five journals teaches us that the weighted impact factor and the SNIP indicator were not significantly correlated in 2007 ($r = 0.75$; *n.s.*). As could be expected (Leydesdorff, 2009), the SJR, the ISI-IF, and the IF for three years are highly correlated among themselves ($p < 0.01$). The weighted impact is correlated with the ISI-IF ($r = 0.90$; $p < 0.05$) and the IF for three years ($r = 0.94$; $p < 0.05$), but not with the SJR ($r = 0.83$; *n.s.*). However, the recently proposed SNIP indicator (Moed, 2010) was not significantly correlated with any of these other indicators.

**Statistics**

Following Kochen (1974) and Pinski & Narin's (1976) "influence weights", weighting citations in terms of citing documents has been proposed more frequently in the literature. Bollen *et al*. (2006), for example, proposed introducing Google's PageRank algorithm (Page *et al*., 1998) into weighting the impact factors. Most recently, Habibzadeh & Yadollahie (2009) proposed a new measure for weighting the impact factors in a study which also contains a review of these proposals. However, following Pinski & Narin (1976), most of these contributions are based on the idea of recursion: the weights are input into an algorithm which converges on an eventual weight. Zitt & Small (2008) proposed the Audience Factor based on the fractional counting, but at the journal level.

Furthermore, these authors also normalized—at the journal level—by dividing averages instead of averaging quotients.

The advantage of our measure is its simplicity and elegance. One obtains distributions of citations because each time a paper is cited, it is cited with a specific weight. These weights are field-specific because they are paper-specific. In other words, no index is needed to determine the field or the subject category (cf. Boyack *et al.*, 2005; Rafols & Leydesdorff, 2009 for the well-known problems of ISI subject categories) because the citing paper positions itself in terms of the relevant fields and thus also in terms of citation behavior. The major advantage, however, is that one obtains full-fledged statistics for testing the distributions of citations for the potential significance of the differences. Note that this can be done for any set, but we focus here on journals and their citation impact.

The three journals studied above, for example, differ *significantly* in terms of their being-cited patterns when Kruskall-Wallis is applied to these distributions (using SPSS v. 15). Additionally, a post-hoc test with Bonferroni correction allows the mean of the differences to be compared between any two journals. The three journals also differ significantly among themselves in terms of this test.

In order to elaborate on this argument, we extended the set with two other mathematics journals which are different in relevant parameters, but belong intellectually to the same field as *Inventiones Mathematicae*. One is *Mathematical Research Letters* which, unlike the journals already examined, contains letters and has a lower impact factor. The other journal is *Annals of Mathematics.* These two journals were selected as most similar on the basis of a factor analysis of the aggregated journal citation environment of *Inventiones Mathematicae* (Leydesdorff, 1987, 2006).

**Multiple Comparisons**

Dependent Variable: weight
Bonferroni

| (I) group | (J) group | Mean Difference (I-J) | Std. Error | Sig. | 95% Confidence Interval | |
|---|---|---|---|---|---|---|
| | | Lower Bound | Upper Bound | Lower Bound | Upper Bound | Lower Bound |
| 1 | 2 | .042295005(*) | .002563890 | .000 | .03509643 | .04949358 |
| | 3 | -.035476788(*) | .003140958 | .000 | -.04429559 | -.02665799 |
| | 4 | .027164987(*) | .004603784 | .000 | .01423905 | .04009093 |
| | 5 | **.007047970** | .003265047 | .309 | -.00211923 | .01621517 |
| 2 | 1 | -.042295005(*) | .002563890 | .000 | -.04949358 | -.03509643 |
| | 3 | -.077771793(*) | .001961675 | .000 | -.08327954 | -.07226404 |
| | 4 | -.015130017(*) | .003895815 | .001 | -.02606821 | -.00419182 |
| | 5 | -.035247035(*) | .002154781 | .000 | -.04129697 | -.02919710 |
| 3 | 1 | .035476788(*) | .003140958 | .000 | .02665799 | .04429559 |
| | 2 | .077771793(*) | .001961675 | .000 | .07226404 | .08327954 |
| | 4 | .062641775(*) | .004297611 | .000 | .05057547 | .07470808 |
| | 5 | .042524758(*) | .002816943 | .000 | .03461569 | .05043383 |
| 4 | 1 | -.027164987(*) | .004603784 | .000 | -.04009093 | -.01423905 |
| | 2 | .015130017(*) | .003895815 | .001 | .00419182 | .02606821 |
| | 3 | -.062641775(*) | .004297611 | .000 | -.07470808 | -.05057547 |
| | 5 | -.020117017(*) | .004389120 | .000 | -.03244025 | -.00779378 |
| 5 | 1 | **-.007047970** | .003265047 | .309 | -.01621517 | .00211923 |
| | 2 | .035247035(*) | .002154781 | .000 | .02919710 | .04129697 |
| | 3 | -.042524758(*) | .002816943 | .000 | -.05043383 | -.03461569 |
| | 4 | .020117017(*) | .004389120 | .000 | .00779378 | .03244025 |

\* The mean difference is significant at the .05 level.

**Table 2**: Bonferroni correction *ex post* on a one-way analysis of variance among the citation patters of the five journals under study.

Table 2 shows that the citation pattern of *Annals of Mathematics* is not significantly different from that of *Inventiones Mathematicae.* These measures are non-parametric, and therefore the division by the number of cited papers—which varies among the different

journals—can also be expected to make a difference. Thus, the citation impacts are tested and not the weighted impact factors. Note that this test provides us with a non-parametric means to organize journals in precise (but not necessarily distinct) groups which are significantly similar in terms of their being cited patterns (Leydesdorff, 2006).

**Conclusions and summary**

We have argued that Moed's (2010) idea of normalizing the citation impact contextually can be elaborated into the weighted citation impact of a set of documents, which can be tested statistically for its significance in relation to other sets, e.g., other scientific journals. This method solves the problem of the field-specificity of citation behavior at the article level, and it also solves the problem of a lack of statistics for comparing the impacts of journals when using the ISI Impact Factors or similar measures. The proposed SNIP indicator, however, does not solve these problems, and can perhaps therefore be reconsidered in favor of using this simpler and more elegant measure.

**Acknowledgment**

The authors wish to acknowledge Koen Frenken and two anonymous referees for contributions and feedback.

**References**

Bollen, J., Rodriquez, M. A., & Van de Sompel, H. (2006). Journal status. *Scientometrics, 69*(3), 669-687.
Boyack, K. W., Klavans, R., & Börner, K. (2005). Mapping the Backbone of Science. *Scientometrics, 64*(3), 351-374.

Cronin, B., & Meho, L. I. (2008). The shifting balance of intellectual trade in information studies. *Journal of the American Society for Information Science and Technology, 59*(4), 551-564.

Egghe, L. (2009). Mathematical derivation of the impact factor distribution. *Journal of Informetrics,* 3(4), 290-295.

Falagas, M. E., Kouranos, V. D., Arencibia-Jorge, R., & Karageorgopoulos, D. E. (2008). Comparison of SCImago journal rank indicator with journal impact factor. *The FASEB Journal, 22*(8), 2623-2628.

Garfield, E. (1972). Citation Analysis as a Tool in Journal Evaluation. *Science* 178(Number 4060), 471-479.

Habibzadeh, F., & Yadollahie, M. (2008). Journal weighted impact factor: A proposal. *Journal of Informetrics,* 2(2), 164-172.

Kochen, M. (1974). Principles of information retrieval. *Melville, Los Angeles, CA*.

Leydesdorff, L. (1987). Various methods for the Mapping of Science. *Scientometrics 11*, 291-320.

Leydesdorff, L. (2006). Can Scientific Journals be Classified in Terms of Aggregated Journal-Journal Citation Relations using the Journal Citation Reports? *Journal of the American Society for Information Science & Technology,* 57(5), 601-613.

Leydesdorff, L. (2008). *Caveats* for the Use of Citation Indicators in Research and Journal Evaluation. *Journal of the American Society for Information Science and Technology,* 59(2), 278-287.

Leydesdorff, L. (2009). How are New Citation-Based Journal Indicators Adding to the Bibliometric Toolbox? *Journal of the American Society for Information Science and Technology, 60*(7), 1327-1336.

Leydesdorff, L., & Opthof, T. (in preparation), Normalization, CWTS indicators, and the Leiden Rankings: Differences in citation behavior at the level of fields, available at http://arxiv.org/abs/1003.3977.

Lundberg, J. (2007). Lifting the crown—citation z-score. *Journal of informetrics, 1*(2), 145-154.

Moed, H. F. (2010). Measuring contextual citation impact of scientific journals. *Journal of Informetrics,* in print.

Moed, H. F., De Bruin, R. E., & Van Leeuwen, T. N. (1995). New bibliometric tools for the assessment of national research performance: Database description, overview of indicators and first applications. *Scientometrics,* 33(3), 381-422.

Narin, F. (1976). *Evaluative Bibliometrics: The Use of Publication and Citation Analysis in the Evaluation of Scientific Activity*. Washington, DC: National Science Foundation.

Opthof, T., & Leydesdorff, L. (2010). Caveats for the journal and field normalizations in the CWTS ("Leiden") evaluations of research performance. *Journal of Informetrics,* in print.

Page, L., Brin, S., Motwani, R., & Winograd, T. (1998). The pagerank citation ranking: Bringing order to the web. available at http://dbpubs.stanford.edu/pub/1999-66: Technical report, Stanford Digital Library Technologies Project, 1998.

Pinski, G., & Narin, F. (1976). Citation Influence for Journal Aggregates of Scientific Publications: Theory, with Application to the Literature of Physics. *Information Processing and Management,* 12(5), 297-312.

Rafols, I., & Leydesdorff, L. (2009). Content-based and Algorithmic Classifications of Journals: Perspectives on the Dynamics of Scientific Communication and Indexer Effects *Journal of the American Society for Information Science and Technology,* 60(9), 1823-1835.

Sher, I. H., & Garfield, E. (1965). New tools for improving and evaluating the effectiveness of research. *Second conference on Research Program Effectiveness, Washinton, DC July*, 27-29.

Small, H. & Sweeney E. (1985). Clustering the Science Citation Index Using Co-Citations I. A Comparison of Methods, *Scientometrics 7,* 391-409.

Van Raan, A. F. J., Van Leeuwen, T. N., Visser, M. S., Van Eck, N. J., & Waltman, L. (2010). Rivals to the crown: reply to Opthof and Leydesdorff. *Journal of Informetrics, 4*(3), forthcoming.

Waltman, L., & van Eck, N. J. (2008). Some comments on the journal weighted impact factor proposed by Habibzadeh and Yadollahie. *Journal of Informetrics,* 2(4), 369-372.

Waltman, L., & Van Eck, N. J. (2009). Some comments on Egghe's derivation of the impact factor distribution. *Journal of Informetrics,* 3(4), 363-366.

Wu, H. (2004). "Order of operations" and other oddities in school mathematics. 3, available at http://math.berkeley.edu/~wu/order5.pdf (Accessed on October 21, 2009.

Zitt, M., & Small, H. (2008). Modifying the journal impact factor by fractional citation weighting: The audience factor. *Journal of the American Society for Information Science and Technology, 59*(11), 1856-1860.